\documentclass[prb,twocolumn,showpacs,preprintnumbers,amsmath,amssymb]{revtex4}
\usepackage{graphicx}
\usepackage{dcolumn}
\usepackage{bm}

\begin{document}

\title{Pressure dependence of the Curie temperature in Ni$_2$MnSn Heusler alloy: A
first-principles study}

\author{E.~\c Sa\c s\i o\~glu, L. M.  Sandratskii and  P. Bruno}

\affiliation{Max-Planck Institut f\"ur Mikrostrukturphysik,
D-06120 Halle, Germany }

\date{\today}

\begin{abstract}
The pressure dependence of electronic structure, exchange
interactions and Curie temperature  in ferromagnetic Heusler alloy
Ni$_2$MnSn has been studied theoretically within the framework of
the density-functional theory. The calculation of the exchange
parameters is based on the frozen--magnon approach. The Curie
temperature, T$_c$, is calculated within the mean--field
approximation by solving the matrix equation for a
multi--sublattice system. In agrement with experiment the Curie
temperature increased from 362K at ambient pressure to 396 at 12
GPa. Extending the variation of the lattice parameter beyond the
range studied experimentally we obtained non--monotonous pressure
dependence of the Curie temperature and metamagnetic transition.
We relate the theoretical dependence of T$_c$ on the lattice
constant to the corresponding dependence predicted by the
empirical interaction curve. The Mn-Ni atomic interchange observed
experimentally is simulated to study its influence on the Curie
temperature.
\end{abstract}

\pacs{75.50.Cc, 75.30.Et, 71.15.Mb, 74.62.Fj}

\maketitle

\section{introduction}

The pressure dependence of the Curie temperature provides
important information on a ferromagnetic system and is an object
of intensive studies both experimental
\cite{expt1,expt2,expt3,expt4} and theoretical.
\cite{theo1,theo2,theo3,theo4,theo5,theo6} The key question here
is the character of the variation of various magnetic properties
with decreasing distances between magnetic atoms. In an early
work, Castellitz \cite{castel} proposed an empirical rule
(interaction curve) that describes the dependence of the Curie
temperature of the Mn-containing ferromagnetic alloys with 4-5
valence electrons per molecule on the ratio $R/d$ where $R$ is the
nearest-neighbor Mn-Mn distance and $d$ is the radius of the
atomic Mn 3d shell. The curve is supposed to represent the Curie
temperatures of various systems at ambient pressure as well as the
pressure dependence of T$_c$ of a given system. The function is
not monotonous and has a maximum at the $R/d$ value of about 3.6
(see Fig.~\ref{fig_interactioncurve}). According to the
interaction curve, one can expect $dT_c/dP>0$ for alloys with
$R/d>3.6$ (e.g., Ni$_2$MnSn and Cu$_2$MnIn). On the other hand,
the systems with $R/d<3.6$ (e.g., NiAs-type MnAs, MnSb and MnBi)
are expected to have negative pressure dependence of the Curie
temperature. These predictions are in agreement with experiment.
\cite{motiv1,motiv2,nias1}

Recently Kanomata \textit {et al.} suggested a generalization of
the interaction curve to the case of 6-7 valence electrons per
chemical formula. \cite{general} These systems form a new branch
of the dependence of the Curie temperature on the Mn-Mn distance
(Fig.~\ref{fig_interactioncurve}). The available experimental
values of the pressure derivative of the Curie temperature,
$dTc/dP$ , for Heusler alloys are consistent with those expected
from the interaction curve.\cite{conf1,conf2,conf3}

Early experiments on the pressure dependence of the Curie
temperature of Heusler alloys have been performed in a low
pressure region (less than 0.5  GPa). Recently Gavriliuk \textit{
et al.}\cite{gavril} have studied structural and magnetic
properties of Ni$_2$MnSn in the pressure interval up to 10.8 GPa.
The authors have found an increasing linear dependence of the
Curie temperature on applied pressure. The M\"{o}ssbaurer
spectroscopy revealed partial interchange of the Mn and Ni atoms.

The purpose of the present work is a first-principles study of the
electronic structure, exchange interactions and Curie temperature
in Ni$_2$MnSn as a function of pressure. The main attention is
devoted to the interval of the interatomic Mn--Mn distances from $4.26 \AA$ to
$4.06 \AA $ that corresponds to the available experimental
variation of this parameter. These values of the Mn--Mn distance
are far above the value of $3.6 \AA$ that, according to
interaction curve, separates the regions of positive and negative
pressure gradients of the Curie temperature for this group of systems.
To verify the appearance of the non-monotonous behavior we extended the
calculation to smaller values of the lattice constant
corresponding to larger applied pressures. We compare empirical
and calculated dependencies. The influence of the Mn-Ni atomic
interchange on the magnetism of the system is also studied.

The  paper is  organized as follows. In Sec. II  we present the
calculational approach. Section III contains the results of the
calculations and discussion. Section IV gives the conclusions.


\section{Calculational Method}

The calculations are carried out with the augmented spherical
waves  method \cite{asw} within the atomic--sphere
approximation.\cite{asa} The exchange--correlation potential is
chosen in the generalized gradient approximation. \cite{gga} A
dense Brillouin zone (BZ) sampling $30\times30\times30$ is used.
To establish the relation between the lattice  parameters and
applied pressure we use the following expression obtained
experimentally in  Ref.\onlinecite{gavril}
\begin{equation}
\frac{(V-V_0)}{V_0}=-aP+bP^2
\end{equation}
where   $a=8.64\cdot10^{-3} GPa^{-1}$,  $b=1.13\cdot10^{-4}
GPa^{-2}$ and $V_0$ is the volume of the unit cell at the ambient
pressure. The radii of all atomic spheres are chosen equal.

\begin{figure}[t]
\begin{center}
\includegraphics[scale=0.32]{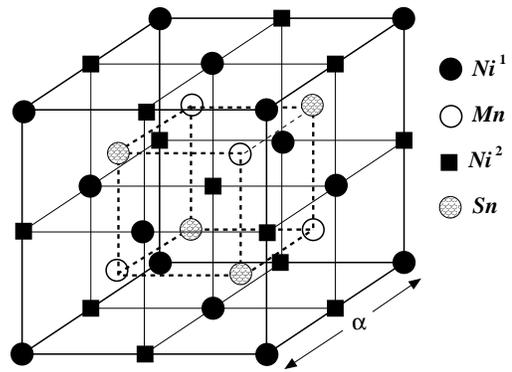}
\end{center}
\caption{ Schematic representation of the $L2_1$ structure adapted
by the  full Heusler alloys. The lattice consists of four
interpenetrating fcc sublattices with the positions $(0,0,0)$ and
$(\frac{1}{2},\frac{1}{2},\frac{1}{2})$ for the Ni and
$(\frac{1}{4},\frac{1}{4},\frac{1}{4})$ and
$(\frac{3}{4},\frac{3}{4},\frac{3}{4})$ for the Mn and Sn,
respectively.}\label{fig_lattice}
\end{figure}

We describe the interatomic exchange interactions in terms of the
classical Heisenberg Hamiltonian
\begin{equation}
\label{eq:hamiltonian2}
 H_{eff}=-  \sum_{\mu,\nu}\sum_{\begin {array}{c}
^{{\bf R},{\bf R'}}\\ ^{(\mu{\bf R} \ne \nu{\bf R'})}\\
\end{array}} J_{{\bf R}{\bf R'}}^{\mu\nu}
{\bf s}_{\bf R}^{\mu}{\bf s}_{\bf R'}^{\nu}
\end{equation}
In Eq.(\ref{eq:hamiltonian2}), the  indices  $\mu$ and $\nu$
number different sublattices and ${\bf R}$ and ${\bf R'}$ are the
lattice vectors specifying the atoms within sublattices, ${\bf
s}_{\bf R}^\mu$ is the unit vector pointing in the direction of
the magnetic moment at site $(\mu,{\bf R})$. The systems
considered contain three 3d atoms in the unit cell with positions
shown in Fig.\ref{fig_lattice}.

We employ the frozen--magnon approach
\cite{magnon_1,magnon_2,magnon_3} to calculate interatomic
Heisenberg exchange parameters. The calculations involve few
steps. In the first step, the exchange parameters between the
atoms of a given sublattice $\mu$ are computed. The calculation is
based on the evaluation of the energy of the frozen--magnon
configurations defined by the following atomic polar and azimuthal
angles
\begin{equation}
\theta_{\bf R}^{\mu}=\theta, \:\: \phi_{\bf R}^{\mu}={\bf q \cdot
R}+\phi^{\mu}. \label{eq_magnon}
\end{equation}
The constant phase $\phi^{\mu}$ is always chosen equal to zero.
The magnetic moments of all other sublattices are kept parallel to
the z axis. Within the Heisenberg model~(\ref{eq:hamiltonian2})
the energy of such configuration takes the form \cite{magnon_2}
\begin{equation}
\label{eq:e_of_q} E^{\mu\mu}(\theta,{\bf
q})=E_0^{\mu\mu}(\theta)+\sin^{2}\theta J^{\mu\mu}({\bf q})
\end{equation}
where $E_0^{\mu\mu}$ does not depend on {\bf q} and the Fourier
transform $J^{\mu\nu}({\bf q})$ is defined by
\begin{equation}
\label{eq:J_q} J^{\mu\nu}({\bf q})=\sum_{\bf R} J_{0{\bf
R}}^{\mu\nu}\:\exp(i{\bf q\cdot R}).
\end{equation}

In the case of $\nu=\mu$ the sum in Eq. (\ref{eq:J_q}) does not
include ${\bf R}=0$. Calculating $ E^{\mu\mu}(\theta,{\bf q})$ for
a regular ${\bf q}$--mesh in the Brillouin zone of the crystal and
performing back Fourier transformation one gets exchange
parameters $J_{0{\bf R}}^{\mu\mu}$ for sublattice $\mu$.

The  determination of the exchange interactions between the atoms of
two different sublattices $\mu$ and $\nu$  is discussed in Ref.
\onlinecite{intersublattice}.

The Curie  temperature is estimated within the mean--field
approximation for a multi--sublattice material by solving the
system of  coupled   equations \cite{intersublattice,Anderson}
\begin{equation}
\label{eq_system} \langle s^{\mu}\rangle
=\frac{2}{3k_BT}\sum_{\nu}J_0^{\mu\nu}\langle s^{\nu}\rangle
\end{equation}
where  $\langle s^{\nu}\rangle$ is the average $z$ component of
${\bf s}_{{\bf R}}^{\nu}$  and $J_0^{\mu\nu}\equiv\sum_{\bf R}
J_{0{\bf R}}^{\mu\nu}$. Eq.(\ref{eq_system}) can be represented in
the form of eigenvalue matrix problem
\begin{equation}
\label{eq_eigenvalue} ({\bf \Theta}-T {\bf I}){\bf S}=0
\end{equation}
where $\Theta_{\mu\nu}=\frac{2}{3k_B}J_0^{\mu\nu}$, ${\bf I}$ is a
unit matrix and ${\bf S}$ is the vector of $\langle s^{\nu}\rangle
$. The largest eigenvalue of matrix $\Theta$ gives the value of
Curie temperature. \cite{Anderson}


\section{results and  discussion}

\begin{table}
\caption{Lattice parameters, magnetic moments and Curie
temperatures in Ni$_2$MnSn at ambient pressure  and applied
pressure  of 16 GPa. For  comparison, the magnetic moment obtained
with full potential FLAPW-GGA method is presented.
\label{lattice}}
\begin{ruledtabular}
\begin{tabular}{lcc}
 &  $a$=6.022 \AA [P=0 GPa] &  $a$=5.821 \AA [P=16 GPa]  \\  \hline
Ni      & 0.21     & 0.19      \\
Mn      & 3.73     & 3.47      \\
Sn      & -0.05    & -0.05     \\
Total   & 4.09,4.10$^{\textmd{a}}$     & 3.81     \\
T$_c$[calc]   &  362,373$^{\textmd{b}}$     & 400      \\
T$_c$[expt]   &  360$^{\textmd{c}}$,  342$^{\textmd{d}}$, 338$^{\textmd{e}}$     &  -      \\

\end{tabular}
\end{ruledtabular}
\begin{flushleft}
$^{\textmd{a}}$Ref.\onlinecite{full-potential}\\
$^{\textmd{b}}$Ref.\onlinecite{Kurtulus}\\
$^{\textmd{c}}$Ref.\onlinecite{webster}\\
$^{\textmd{d}}$Ref.\onlinecite{gavril}\\
$^{\textmd{e}}$Ref.\onlinecite{kyu}\\
\end{flushleft}
\end{table}

We will subdivide the discussion of the influence of the pressure
on the electronic properties of Ni$_2$MnSn in two parts. First, we
present a detailed study of the low-pressure region where an
experimental information is available (We extend this interval up
to $\sim$20 GPa). In particular we verify the monotonous increase
of the Curie temperature with increasing pressure in this region.
Then we consider a much larger interval of the variation of the
lattice parameter to study the occurrence of the non--monotonous
behavior of the Curie temperature.

\subsection{Low pressure region}

\begin{figure}[t]
\begin{center}
\includegraphics[scale=0.44]{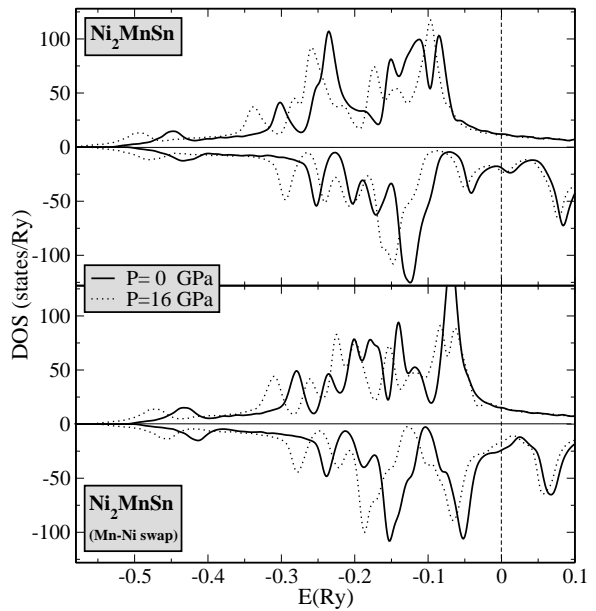}
\end{center}
\caption{Upper panel: Spin projected density of  states of
Ni$_2$MnSn  for ambient pressure   and applied pressure  of 16
GPa. Lower panel: The same for the case of  Mn-Ni atomic interchange.}
\label{fig_dos}
\end{figure}

\begin{figure}[t]
\begin{center}
\includegraphics[scale=0.43]{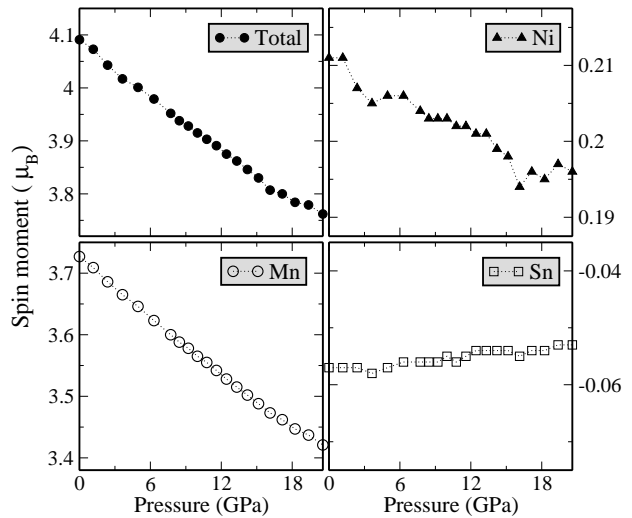}
\end{center}
\caption{Pressure dependence of magnetic moments in Ni$_2$MnSn.}
\label{fig_moments}
\end{figure}

We  begin with the discussion of the effect of  pressure on the
electronic structure. In Fig.~\ref{fig_dos}, we compare the
density of states calculated for the ambient pressure and the
applied pressure of 16 GPa. As expected, the pressure leads to the
broadening of the bands that stems from the decreasing
inter-atomic distances  and, therefore, increasing overlap of the
atomic states. One of the consequences of the band broadening is
the trend to the decrease of the magnetic moments. This trend is
demonstrated in table~\ref{lattice} and Fig.~\ref{fig_moments}. In
Fig.~\ref{fig_moments} we present a detailed information on the
atomic and total magnetic moments for the range of pressures up to
20.6 GPa. The dependence of the Mn magnetic moment on pressure can
be well represented by a linear function with a negative slope.
The behavior of the induced moment of Ni is more peculiar. The
dependence deviates strongly from the straight line and shows weak
oscillations in the high-pressure part of the curve. This behavior
reflects the details of the pressure dependence of the band
structure, in particular, the form of the DOS in the energy region
close to the Fermi level and the character of the Mn-Ni
hybridization. Since these weak oscillations do not play
noticeable role in the issues we focus on in this paper we do not
further investigate their origin. The induced moment on Sn has the
direction opposite to the direction of the Mn moment. Its value
decreases slowly with increasing pressure. The spin polarization
at the Fermi level shows very weak pressure dependence.

Thus the decreasing lattice constant produces a clear trend to a
monotonous decrease of the atomic magnetic moments. For our
purpose of the investigation of the pressure dependence of the
Curie temperature it is important to relate the increasing band
width and decreasing magnetic moments to the properties of the
inter-atomic exchange interactions.

In the spirit of the Heisenberg model of localized moments one
expects that decreasing atomic moments produce the trend to the
decrease of the inter-atomic exchange interactions by the factor
of $M_p^2/M_0^2$ where $M_p$ is the atomic moment at pressure $P$
and $M_0$ is the moment at the ambient pressure. Correspondingly,
one expects the trend to decreasing Curie temperature resulting
from decreasing atomic moments.

An opposite monotonous trend to increasing interatomic  exchange
interactions is produced by increasing electron hopping and, as a
result, more efficient mediation of the exchange interactions
between magnetic atoms. The competition of two opposite trends
opens possibility for both increase and decrease of the Curie
temperature with applied pressure as well as for a nonmonotonous
pressure dependence in a larger pressure interval.

\begin{figure}[t]
\begin{center}
\includegraphics[scale=0.42]{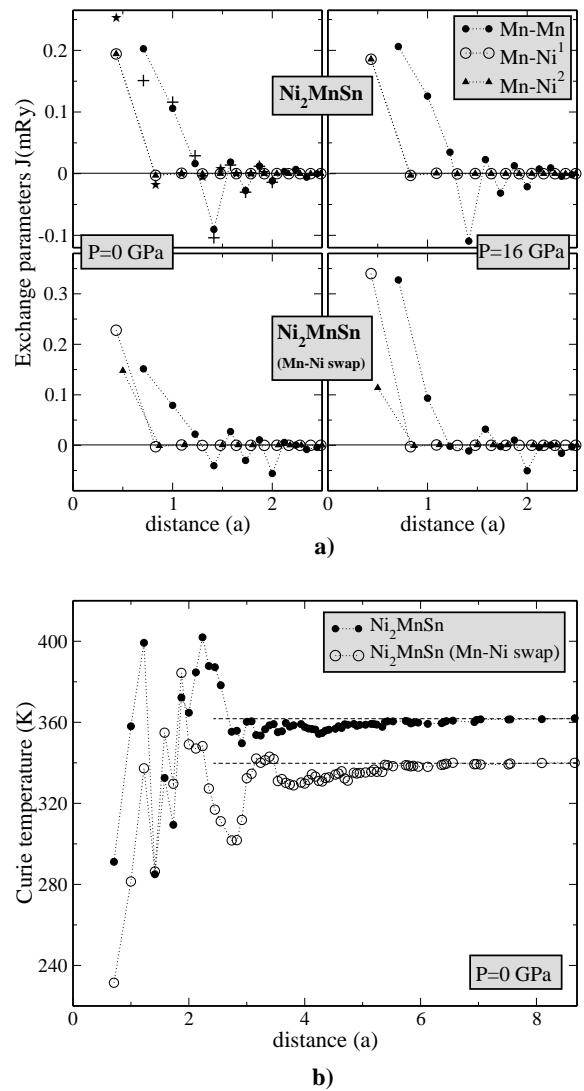}
\end{center}
\caption{ a)Upper panel: Interatomic exchange parameters of
Ni$_2$MnSn for ambient pressure and applied  pressure of 16 GPa.
Lower panel: The same for the case of  Mn-Ni atomic interchange.
Zero pressure comparison of  both Mn--Mn (+) and  Mn--Ni ($\star$)
exchange parameters with Ref.\onlinecite{Kurtulus} is  also shown.
b) Pressure variation of the Curie temperature  with increasing
number of coordination spheres  with and  without Mn-Ni atomic
interchange.} \label{fig_exchange}
\end{figure}

In Fig. \ref{fig_exchange}a,  we present the calculated
inter-atomic exchange parameters of Ni$_2$MnSn for pressures of 0
and 16 GPa.  For comparison, a zero-pressure result  of previous
calculation is also presented.  At both pressures the patterns of
inter-atomic exchange interactions are very similar. This
similarity involves both Mn--Mn and Mn--Ni exchange interactions.
The Mn--Mn interactions are long-ranged reaching beyond the 8th
nearest neighborhood distance and have the RKKY-type oscillating
character. The inter-sublattice Mn--Ni interaction behaves very
differently. A sizable interaction takes place only between
nearest neighbors. Note  that  Fig.\ref{fig_exchange}a does not
present all calculated exchange parameters: the exchange
parameters have been evaluated up to the inter-atomic distances of
$8.7a$ that corresponds to about 70 coordination spheres. The
absolute value of the parameters decays quickly with increasing
interatomic distance. In Fig.~\ref{fig_exchange}b, we demonstrate
the convergence of the calculated Curie temperature with respect
to increasing number of the atomic coordination spheres. The main
contribution to T$_c$ comes from the interaction between atoms
lying closer than $3a$. After $5a$ no sizable contribution is
detected.

\begin{figure}[t]
\begin{center}
\includegraphics[scale=0.36]{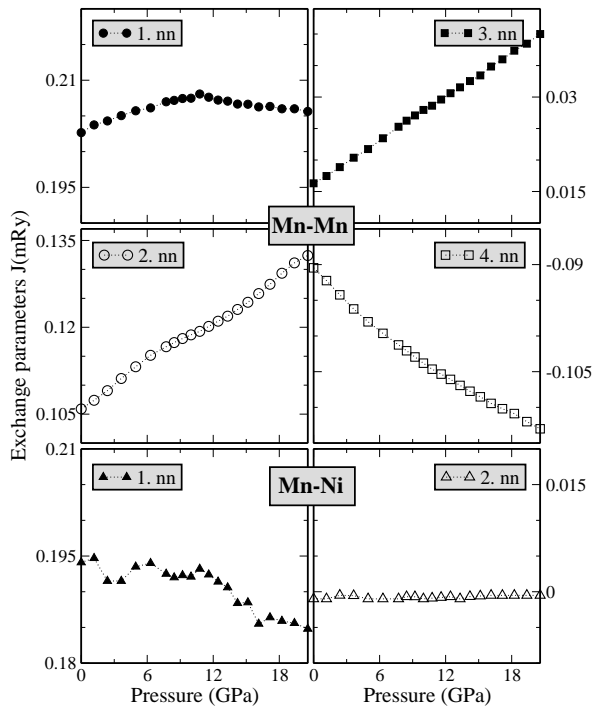}
\end{center}
\caption{Pressure dependence of the first four nearest neighbor
Mn--Mn exchange parameters and two Mn--Ni exchange parameters.
\label{fig_dominantexchange}}
\end{figure}

In  Fig.\ref{fig_exchange}a, we compare our exchange parameters
for  Ni$_2$MnSn at zero pressure with the exchange parameters
calculated recently by Kurtulus \textit{ et al.} \cite{Kurtulus}
Kurtulus \textit{et al.} used TB-LMTO-ASA method  and local
spin-density approximation (LSDA). The inter-atomic exchange
parameters were evaluated using the real-space approach by
Liechtenstein \textit{ et al.} \cite{real-space} This approach and
the frozen-magnon technique employed in the present paper are
equivalent to each other. In the real-space method by
Liechtenstein \textit{ et al.} the inter-atomic exchange
parameters are calculated directly whereas in the frozen-magnon
approach they are obtained by the Fourier transformation of the
magnon dispersion.

In Table \ref{lattice}, we  present the MFA estimation of the
Curie temperature. It is in good agreement with available experimental
values overestimating them somewhat.
An overestimation of the Curie temperature is a usual
feature of the MFA.\cite{sabiryanov1,sabiryanov2,pajda} It arises
from the property that the MFA expression for  T$_c$ corresponds
to an equal weighting of the low- and high--energy spin--wave
excitations. A better weighting of the magnetic excitations is
provided by the random--phase approximation (RPA). \cite{pajda}
However, in the case of the lattices with high atomic coordination
numbers and in the cases of the magnon dispersion deviating
strongly from a simple sinusoidal form the both MFA and RPA can
give similar values of the Curie temperature making the MFA
estimation reliable. \cite{zb,bouzerar} A good agreement of the
theoretical and experimental T$_c$ values shows that the MFA is
well applicable for the given system.

\begin{figure}[t]
\begin{center}
\includegraphics[scale=0.41]{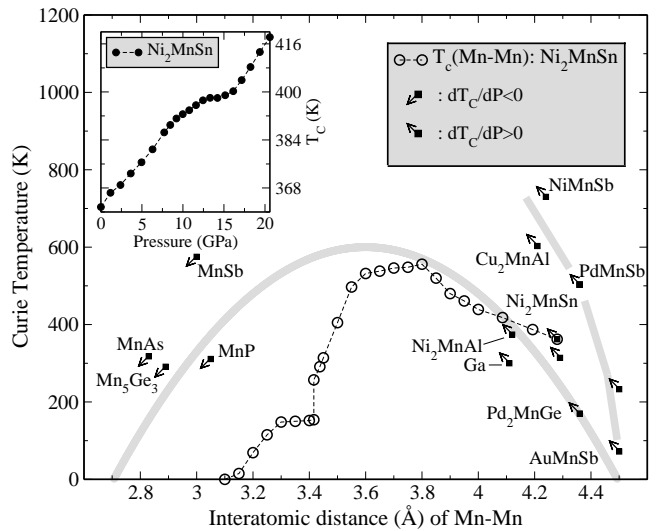}
\end{center}
\caption{Schematic representation  of Kanomata's empirical
interaction curve and the calculated Curie temperature  as a
function of the interatomic Mn--Mn distance in Ni$_2$MnSn. Inset
shows pressure variation of T$_c$ in the low pressure region.
Small  rectangles present the Curie temperatures of the
corresponding compounds at ambient pressure. The attached arrows
show the sign of $dT_c/dP$. (The experimental information is taken
from Ref.\onlinecite{webster}).} \label{fig_interactioncurve}
\end{figure}

The pressure dependence of the interatomic exchange parameters is
presented in Fig.~\ref{fig_dominantexchange}. The corresponding
Curie temperature is shown in inset in
Fig~\ref{fig_interactioncurve}. The analysis shows that the
leading contribution into Curie temperature is given by the Mn--Mn
exchange interactions within the first three coordinations
spheres. The numbers of the atoms in these spheres are 12, 6 and
24, respectively for the first, second and third spheres. The
exchange parameters corresponding to the second and third
coordination spheres increase monotonously with increasing
pressure determining the increase of the Curie temperature
(Fig~\ref{fig_interactioncurve}). Thus, the increase of the
experimental Curie temperature with pressure in the corresponding
pressure region \cite{gavril} is well confirmed by the
calculations. In terms of the competition of the two opposite
trends discussed above this result means a stronger effect of the
increasing hopping compared with the effect of decreasing atomic
moments. Such a behavior is expected for large inter-atomic
distances.

Note that in the interval from 9 GPa to 16 GPa we obtain a flat
feature in the pressure dependence of the Curie temperature. This
behavior is in a correlation with the recent experiment of Kyuji
\textit{ et al.}\cite{kyu} who obtained the Curie temperature
increase  from 338K at the ambient pressure to 395K at 12 GPa. At
$\sim$7 GPa they obtained a decrease of the pressure gradient that
can be put into correspondence to the theoretical flat feature.

Both measured and calculated Curie temperatures are in good
agreement with the empirical interaction curve for the
corresponding region of the Mn-Mn distances
(Fig.~\ref{fig_interactioncurve}).

\subsection{High pressure region}

\begin{figure}[t]
\begin{center}
\includegraphics[scale=0.42]{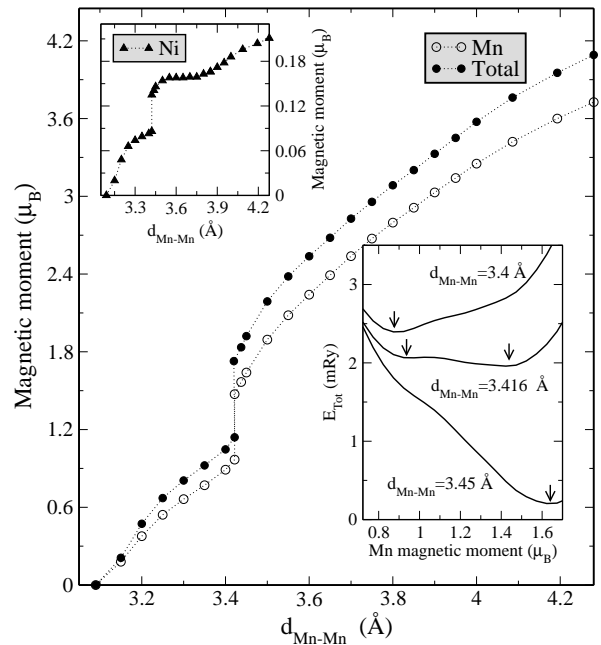}
\end{center}
\caption{Mn and total magnetic moments as a function of the Mn--Mn
distance. Upper inset shows the variation of induced Ni moment
with Mn--Mn distance. Lower inset shows E$_{Tot}$ as a function of
Mn magnetic moment for selected Mn--Mn interatomic distances.
Arrows indicate the energy minima.} \label{fig_fixmoment}
\end{figure}

To verify the non-monotonous pressure dependence of $T_c$
predicted by the interaction curve we extended the calculation to
smaller Mn-Mn distances down to $3.09\AA$. The calculated magnetic
moments are  presented in Fig.~\ref{fig_fixmoment}. The Mn moment
decreases with the reduction of Mn-Mn distance. An interesting
feature is obtained at $d_{Mn-Mn}=3.416 \AA$ where the value of
the magnetic moment changes discontinuously. To study the origin
of the discontinuity we employed the fixed--spin--moment method
\cite{fix-moment-1,fix-moment-2,fix-moment-3} that allows the
calculation of the total energy as a function of the spin moment
for a given lattice parameter. The corresponding curves are
presented in the inset in Fig.~\ref{fig_fixmoment}. For large
Mn--Mn distance the curve has one minimum corresponding to a
high--spin state. In the region of the discontinuity the curve has
two local minima revealing the presence of a metastable state. At
the point of discontinuity the minimum corresponding to the
low--spin state becomes lower. With further decrease of the
lattice volume the minimum corresponding to the high-spin state
disappears. At $d_{Mn-Mn}=3.09 \AA$ the magnetic moment vanishes
and the ground state of the system becomes a Pauli--paramagnet.

\begin{figure}[t]
\begin{center}
\includegraphics[scale=0.46]{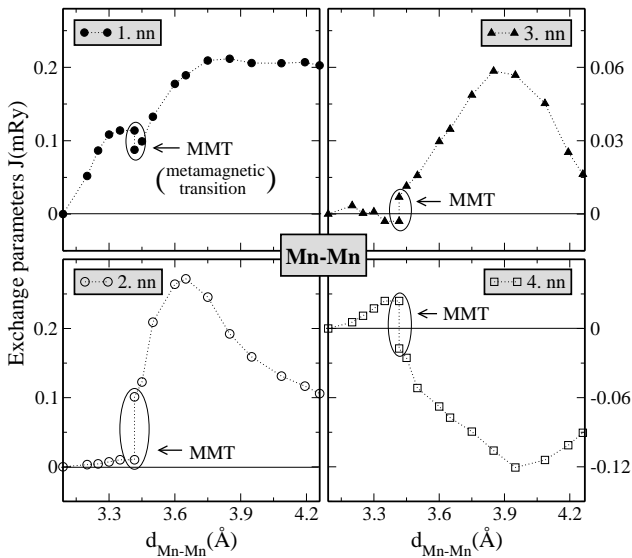}
\end{center}
\caption{The first four nearest neighbor Mn-Mn exchange
interactions in Ni$_2$MnSn as  function of  Mn-Mn interatomic
distance. The ellipses show the region of the metamagnetic transition.}
\label{fig_dominantexchange-Mn-Mn}
\end{figure}

In Fig.~\ref{fig_dominantexchange-Mn-Mn}, the first four nearest
neighbor Mn-Mn exchange  interactions are presented for the broad
interval of the Mn--Mn  interatomic distance. The pressure region
discussed in the preceding section corresponds to the last three
points in  the  plot. Three of the four leading parameters show
nonmonotonous behavior that is reflected in the nonmonotonous
behavior of the Curie temperature
(Fig.~\ref{fig_interactioncurve}). The absolute values of the 2nd,
3rd and 4th neighbor Mn--Mn parameters first increase with
pressure, reach their maxima at the Mn--Mn distances in the region
from about $4.0 \AA$ to about $3.6 \AA$ and decrease strongly with
further decrease of the Mn--Mn distance. There are some weak
peculiarities in the behavior of the exchange parameters at the
region after the discontinues transition such as additional local
extrema in the 1st, 3th and 4th neighbor parameters. They,
however, compensate each other and  the Curie temperature has only
one maximum at the Mn--Mn distance of about 3.8 $\AA$
(Fig.\ref{fig_interactioncurve}).

The non--monotonous behavior of the exchange parameters and T$_c$
can be interpreted as a result of the competition of two opposite
monotonous trends appearing with the variation of the Mn--Mn
distances discussed in the previous section. In the low--pressure
region the influence of the increasing hopping prevails while in
the high--pressure region the influence of decreasing magnetic
moments becomes more important.

Qualitatively, the calculated pressure dependence of T$_c$ in the
broad pressure interval is in agreement with the Kanomata's
empirical interaction curve. Indeed, we obtained non--monotonous
pressure dependence characterized by one maximum separating the
regions of positive and negative pressure gradients. The
low--pressure part of the calculated dependence is in reasonable
quantitative agreement with the Kanomata's  interaction curve. For
smaller lattice volumes the calculated Curie temperature decreases
faster than it is prognosticated by the interaction curve. The
calculations predict the discontinuity in the pressure dependence
of the Curie temperature of Ni$_2$MnSn which is absent in the
empirical interaction curve. The extension of the measurements to
higher pressures is desirable to verify the predictions of the
calculations.

\subsection{Atomic inter-sublattice interchange}

The calculations of the Curie temperature of Ni$_2$MnSn  discussed
in the preceding sections are in good correlation with measured
$T_c$ values for the range of pressures studied experimentally. A
detailed numerical comparison shows, however, that the theoretical
pressure derivative, $dT_c/dP$, estimated as 3.22 K/GPa is
substantially smaller than the experimental estimation of
7.44K/GPa obtained  by Gavriliuk \textit{ et al.}\cite{gavril} To
verify the role of the atomic interchange between Ni and Mn
sublattices observed experimentally \cite{gavril} we performed
calculation for a model system where the atoms of the Mn
sublattice are interchanged with the atoms of one of the Ni
sublattices. Although this model is a strong simplification of the
experimental situation it allows the investigation of the trends
resulting from the Mn--Ni interchange.

With Mn--Ni interchange  we obtain  a substantial difference in
the electron structure  of the system. The corresponding DOS and
magnetic moments for ambient and applied pressure  of 16 GPa  are
presented in Fig.\ref{fig_dos} and table \ref{lattice-swap}. In
this case the Mn states hybridize differently with the states of
two Ni atoms. As a result the magnetic moments of the Ni atoms
assume different values. The total magnetic moment per formula
unit decreases from 3.50$\mu_B$ at ambient pressure to the
3.09$\mu_B$ at the applied pressure of 16 GPa. The decrease of the
total magnetic moment is mostly the result of the reduction of the
Mn moment. Note that at the pressure of 16 GPa the relative
variation of the total magnetic moment is two times larger than in
Ni$_2$MnSn without Mn--Ni interchange. The change in the shape of
the 3d peaks and broadening of the bands are similar to those for
the system without swapping.

\begin{table}
\caption{Lattice parameters, magnetic moments and Curie
temperatures at ambient pressure and applied pressure  of 16 GPa
for Ni$_2$MnSn with  Mn--Ni atomic
interchange.\label{lattice-swap}}
\begin{ruledtabular}
\begin{tabular}{lcc}
 &  $a$=6.022 \AA[P=0 GPa] &  $a$=5.821 \AA[P=16 GPa]\\
 \hline
           Ni$^1$   & 0.21  & 0.26      \\
           Ni$^2$  & 0.08   & 0.05      \\
           Mn      & 3.24   & 2.80      \\
           Sn      & -0.04  & -0.03     \\
           Total   &  3.50  & 3.09      \\
           T$_c$   &  340   &  562      \\
\end{tabular}
\end{ruledtabular}
\end{table}

At zero pressure the  pattern of  exchange parameters
(Fig.~\ref{fig_exchange}) and resulting Curie temperature (Table
\ref{lattice-swap}) are very similar to the case without Mn--Ni
interchange. However, the  situation  is different at applied
pressure of 16 GPa. Both  Mn-Mn and Mn-Ni$^1$ nearest-neighbor
exchange parameters increase substantially. The remaining exchange
parameters show small pressure dependencies. The interaction  of
the Mn moment with the moment of the second  Ni atom is slightly
reduced.  The substantial increase of the leading exchange
parameters with pressure  results in considerable  change of the
Curie temperature from 400 K at ambient pressure  to 562 K  at 16
GPa.  Assuming a linear variation of  T$_c$ with  pressure we
estimate the pressure derivative, $dT_c/dP$,  as 12.5 K/GPa. This
value of $dT_c/dP$ exceeds strongly the corresponding value for
the system without Mn--Ni atomic interchange. Since the number of
the swaped Mn and Ni atoms in our model is much larger than in the
samples measured the calculated $dT_c/dP$ cannot be directly
compared with the experimental pressure derivative. Important,
however, that the Mn--Ni atomic interchange increases the pressure
derivative of T$_c$ that gives an explanation for the low value of
the theoretical pressure derivative in the case of the system
without swapping. A detailed study of the influence of the
inter--sublattice atomic interchange on the electron properties of
the Heusler systems is an interesting extension of the present
study.

\section{conclusion}

In conclusion, we  have  systematically  studied the pressure
dependence of  exchange interactions and Curie temperature in full
Heusler alloy Ni$_2$MnSn within the parameter--free density
functional theory. We show that the character of the pressure
dependence of the exchange  interactions is a consequence of the
complex interplay of competing trends in the electronic properties
of the system. In agreement with experiment, the Curie temperature
increases with increasing pressure in the pressure region studied.
Extending our theoretical study onto a larger pressure interval we
obtained non--monotonous T$_c$ dependence and the presence of the
metamagnetic transition. The T$_c$ behavior in the whole pressure
interval is in qualitative correlation with the Kanomata's
empirical interaction curve. In the low--pressure region there is
good quantitative agreement between calculated values and the
prediction of the empirical rule. The Mn--Ni atomic interchange is
shown to increase the pressure derivative of the Curie temperature
that suggests the physical mechanism for the improved agreement
between the experimental and theoretical estimations of this
parameter.

\begin{acknowledgments}
The financial support of Bundesministerium f\"ur Bildung und
Forschung is acknowledged.
\end{acknowledgments}

\end{document}